# Near-infrared to visible upconversion imaging using a broadband pump laser

ROMAIN DEMUR,[1,2,*] RENAUD GARIOUD,[1] ARNAUD GRISARD,[1] ERIC LALLIER,[1] LUC LEVIANDIER,[1] LOÏC MORVAN,[1] NICOLAS TREPS,[2] AND CLAUDE FABRE[2]

[1]*Thales Research & Technology, 1 Avenue Augustin Fresnel, 91767 Palaiseau, France*
[2]*Laboratoire Kastler Brossel, Sorbonne Université, CNRS, ENS-University PSL, Collège de France; 4 Place Jussieu, 75252 Paris, France*
*\*romain.demur@thalesgroup.com*

**Abstract:** We present an upconversion imaging experiment from the near-infrared to the visible spectrum. Using a dedicated broadband pump laser to increase the number of resolved elements converted in the image we obtain up to 56x64 spatial elements with a 2.7 nm wide pump spectrum, more than 10 times the number of elements accessible with a narrowband laser. Results in terms of field of view, resolution and conversion efficiency are in good agreement with simulations. The computed sensitivity of our experiment favorably compares with direct InGaAs camera detection.

## 1. Introduction

Optical systems using near-infrared (NIR) illumination lasers are widely used for remote sensing and LIDAR applications [1–3]. In particular, an illumination laser wavelength around 1.5 μm can be used to benefit from the advantages of the NIR spectral domain, a good transparency range of the atmosphere, the eye safety property and the availability of high peak power lasers. However, detectors in this spectral domain, commonly based on InGaAs materials, suffer from higher readout noise, lower speed and more stringent cooling requirements than their visible counterpart, based on silicon [4], thus limiting the imaging range.

Image frequency upconversion from near-infrared to visible wavelengths by nonlinear parametric sum-frequency mixing in a $\chi^{(2)}$ medium allows to use low noise CMOS or CCD cameras to detect a 1.5 μm signal with enhanced detection performances. This idea, as old as non-linear optics, suffered for a long time from low efficiency non-linear crystals [5–7]. The advent of crystals with high non-linear coefficients such as periodically-poled lithium niobate (PPLN) has brought a renewed interest to upconversion detection [3, 8, 9].

Nevertheless, there are several remaining limitations. Energy conservation given by:

$$\omega_s + \omega_p = \omega_c \qquad (1)$$

where s stands for the signal, p for the pump and c for the converted beam and phase matching condition impose a specific wavelength for the signal and a small acceptance in term of converted spatial modes. Indeed the phase matching condition corresponding to momentum conservation is written in quasi-phase matching configuration as:

$$\overrightarrow{\Delta k} = \vec{k}_s + \vec{k}_p - \vec{k}_c + \vec{k}_{QPM} = \vec{0} \qquad (2)$$

where $k_{QPM} = \frac{2\pi}{\Lambda}$ and $\Lambda$ is the $\chi^{(2)}$ inversion period of the crystal. This condition is fulfilled for a well-defined signal direction. For a signal coming with a different angle, the conversion efficiency drops dramatically following:

$$\eta \propto \text{sinc}^2\left(\frac{\Delta k_z L}{2}\right) \qquad (3)$$

where L is the length of the nonlinear crystal along the propagation direction and $\Delta k_z$ is the corresponding component of the phase mismatch [10].

This condition is a strong limitation to image upconversion because it restricts in practice the number of spatial modes that can be converted. Several refinements have been proposed to increase this number. Some authors used different signal wavelengths to satisfy the phase matching condition for different incoming signal angles: a dual illumination wavelength [11], an ASE illumination source [12] or a supercontinuum source [13]. However, they converted only a small spectral part of their signal in a given direction, thus letting limited opportunities to increase the detection sensitivity. In [14] authors used a temperature gradient inside the crystal to obtain a gradient of phase matching conditions, but with a deleterious reduction on the conversion efficiency. In [15] this limitation is used as an advantage to perform hyperspectral imaging, using incoherent light as a source, but once again with low conversion efficiency.

Conversely, many upconversion experiments have been proposed to obtain high conversion efficiency, but without consideration of spatial multimode signals. Guided wave configurations have been widely used for optical communication [9, 16] where authors achieved up to 90% conversion efficiency for a moderate pump power. Other groups used this technique to perform spectroscopy [3] and LIDAR applications [2].

In order to increase sensitivity in an upconversion imaging system compared to a direct InGaAs detection and obtain images with a sufficient number of resolved spatial elements, both significant conversion efficiency and high angular acceptance of the conversion are needed. To this end, working on the pump properties may help to solve the problem. Changing the pump wavelength changes the phase matching condition and a spatially complex signal with non-collinear phase matching can be converted. Particularly, the use of a pump laser with a few nm wide spectrum should significantly increase the field of view of the upconversion imaging system. The peak power needed to achieve high conversion efficiency is easily obtained using a pulsed pump. In so doing, we expect to obtain a high definition image conversion with good conversion efficiency.

In what follows, we investigate the use of a broadband pump laser to enlarge the upconverted images. We introduce image simulation results from a beam propagation code to show the advantages of using a broadband pump laser compared to a narrowband one. We then present our upconversion experiment using such a broadband pump laser. Results in terms of field of view and number of spatial elements resolved for different pump spectra are discussed. We finally highlight conversion efficiency results with a special emphasis on sensitivity improvement compared to direct InGaAs camera detection.

## 2. Non-collinear upconversion modelling

Non-linear conversion is strongly affected by the phase matching condition. This section aims at developing equations and simulations to understand the non-collinear upconversion of a spatially complex signal with a broadband pump spectrum.

### 2.1 Theoretical analysis

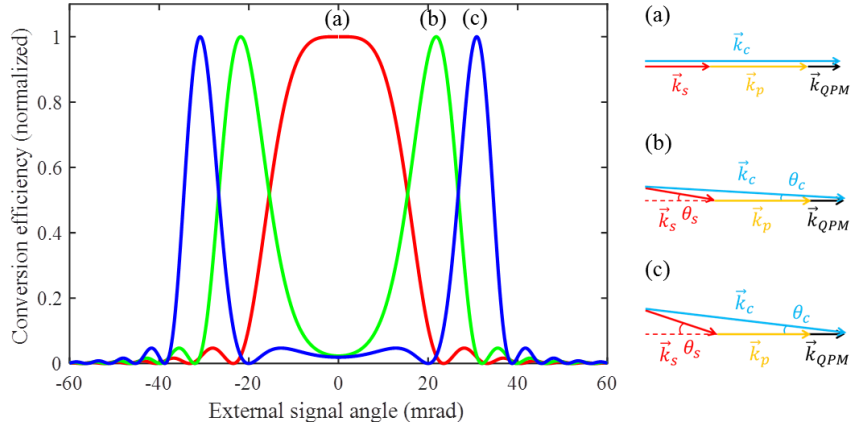

Fig. 1. Normalized conversion efficiency as a function of the incident signal angle at $\lambda_s$=1563 nm for different pump wavelengths. (a) $\lambda_p$=1063 nm, (b) $\lambda_p$=1063.36 nm, (c) $\lambda_p$=1063.72 nm and sketch of the corresponding quasi-phase matching conditions to the right to the maximum of conversion efficiency.

The phase matching condition along the optical axis can be expressed as a function of the pump wavelength change and the signal angle $\theta_s$ inside the crystal. Assuming small signal angles inside the crystal, exactly collimated pump beam, and small spectral broadening of the pump, the phase mismatch expression can be expressed easily form development of the phase matching condition Eq. (2) as:

$$\Delta k_z = \delta k_p - \delta k_c + \frac{k_s}{2}\left(\frac{k_s}{k_c} - 1\right)\theta_s^2 \qquad (4)$$

where $\delta k_p$ and $\delta k_c$ are the wave vector variations along the optical axis. Fig. 1 shows the normalized conversion efficiency as a function of the incoming signal angle outside the crystal for different pump wavelengths in a 2 cm long PPLN crystal plotted with equations (3) and (4). The change in phase matching condition changes the incoming signal angle that is converted with maximum conversion efficiency.

The converted angle can be analytically calculated as a function of pump wavelength change. Considering a small pump spectrum broadening, the pump wavelength can be written as: $\lambda_p + \delta\lambda_p$, with $\lambda_p$ the pump wavelength corresponding to collinear phase matching and $\delta\lambda_p/\lambda_p \ll 1$. The converted wavelength is defined in the same way as: $\lambda_c + \delta\lambda_c$. As a consequence, we can perform Taylor expansion on the Sellmeier expression of the PPLN refractive index: $n_i + \delta n_i = n_i(1 + \alpha_i\, \delta\lambda_i/\lambda_i)$, where $i = p, c$. For a PPLN temperature of T = 97°C, we obtain $\alpha_p$ = -0.0262 and $\alpha_c$ = -0.0674. The maximally converted external signal angle is finally expressed as:

$$\theta_{s,ext} = \frac{\lambda_s}{\lambda_p} \sqrt{2 \frac{n_c(1-\alpha_c) - n_p(1-\alpha_p)}{\frac{\lambda_s}{n_s} - \frac{\lambda_p}{n_p}} \delta\lambda_p} \tag{5}$$

We see on Eq. (5) that the pump wavelength needs to be increased in order to increase the angular signal converted. It is also observed that the angular field of view increase as the square root of the pump spectral broadening $\delta\lambda_p$.

For a given couple of signal angle and pump wavelength following the phase matching condition, the pump wavelength acceptance of the conversion can be calculated. One deduces the maximum pump wavelength shift necessary to have an overlap at half maximum between two adjacent sectors of angular acceptance. This shift is:

$$2\Delta\lambda_p = \frac{2.78}{\pi L} \frac{\lambda_p^2}{n_c(1-\alpha_c) - n_p(1-\alpha_p)} \tag{6}$$

The numerical application gives a shift of 0.36 nm, consistent with the observation on the Fig. 1 where there is a correct overlap between curves (a), (b) and (c) corresponding to pump wavelengths separated by 0.36 nm. This equation is also independent of the signal angle converted. The consequence of Eqs. (5) and (6) is that the pump power necessary to convert a broad image evolves as the square of the maximum signal angle, hence as the solid angle field of view of the imager.

## 2.2 Imaging simulation results

To anticipate the improvements brought by our upconversion setup, we developed a numerical tool to simulate the propagation of the three interacting waves in the imaging device. The propagation through the crystal is governed by the following equations [10]:

$$\frac{\partial E_c}{\partial z} = \frac{2i\omega_c^2 d_{eff}}{k_c c^2} E_s E_p\, e^{-i\Delta kz} + \frac{i}{2k_c}\left(\frac{\partial^2 E_c}{\partial x^2} + \frac{\partial^2 E_c}{\partial y^2}\right) \tag{7}$$

$$\frac{\partial E_s}{\partial z} = \frac{2i\omega_s^2 d_{eff}}{k_s c^2} E_c E_p^* e^{i\Delta kz} + \frac{i}{2k_s}\left(\frac{\partial^2 E_s}{\partial x^2} + \frac{\partial^2 E_s}{\partial y^2}\right) \tag{8}$$

$$\frac{\partial E_p}{\partial z} = \frac{2i\omega_p^2 d_{eff}}{k_p c^2} E_c E_s^* e^{i\Delta kz} + \frac{i}{2k_p}\left(\frac{\partial^2 E_p}{\partial x^2} + \frac{\partial^2 E_p}{\partial y^2}\right) \tag{9}$$

These equations are numerically integrated using a classical symmetrized split-step Fourier method [17] where we split the linear and the nonlinear terms in the equations, considering that the diffraction and the nonlinear interaction act independently over a small propagation distance.

The linear term, corresponding to diffraction – i.e. the second derivative terms in Eqs. (7) – is solved in Fourier space using the Fresnel propagator [18] whereas the nonlinear term – corresponding to the coupled terms in the equations – is solved in direct space using implicit finite difference integration scheme. This numerical method has been widely used in optical fiber propagation problems [19] and to model optical interaction in nonlinear media [20] thanks to its short run time compared to other methods such as finite difference.

The simulation code takes into account wave diffraction, nonlinear interaction with phase mismatch and physical crystal aperture. The code also considers the optical arrangement before and after the crystal. In our configuration, the crystal center is positioned in the Fourier plane of a 4-f optical system. These equations are solved for one specified couple of pump and signal wavelengths. To generalize it to the broadband pump laser, the pump laser emission spectrum is discretized with a distribution of discrete frequencies corresponding to adjacent sectors of angular acceptance defined by Eq. (6). Due to the cardinal sine form of the conversion efficiency the adjacent sectors overlap. In principle mixed conversion between a discretized frequency pump and the upconverted field of an adjacent angular sector should give rise to new fields at a shifted signal frequencies. Such conversions are not considered in the model as it should not affect images quality. In the same way, each angular sector of the signal field are considered to interact only with the pump field having the corresponding discretized frequency. This implies to run the code independently for different pump wavelengths and the upconverted image is simply the sum of the images obtained at these different wavelengths. As a consequence, simulations results for the broadband pump are used to study images definition.

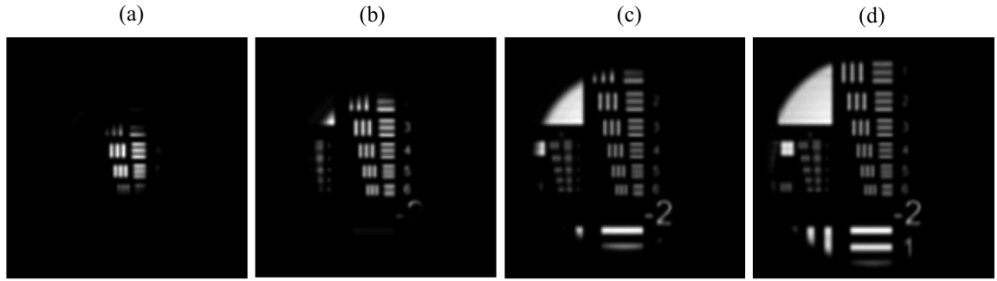

Fig. 2. Image simulation results using the pattern of a 1951 USAF resolution target (a) narrowband pump spectra, (b) 1 nm wide pump spectra, (c) 2 nm wide spectra, (d) 2.7 nm wide spectra.

Fig. 2 shows simulation results of upconverted images for different pump spectra, corresponding to spectra we can achieve experimentally in further sections. Images exhibit a manifest improvement in image size, the broader the pump spectrum, the broader is the field of view.

## 3. Experimental setup and images results

The layout of the experiment is described in Fig. 3. A continuous wave fiber laser at 1563 nm with a power of 1 mW is collimated and expanded to coherently propagate through a test pattern. Two different pump lasers at 1064 nm are successively used to achieve upconversion. Pump and signal are combined with a dichroic mirror (HT@1064 nm, HR@1563 nm). Sum frequency is performed by means of a 20 mm long 5% MgO-doped PPLN crystal with a $d_{eff} \sim 15$ pm/V nonlinear coefficient and a $\Lambda = 11.7$ μm poling period and an aperture of

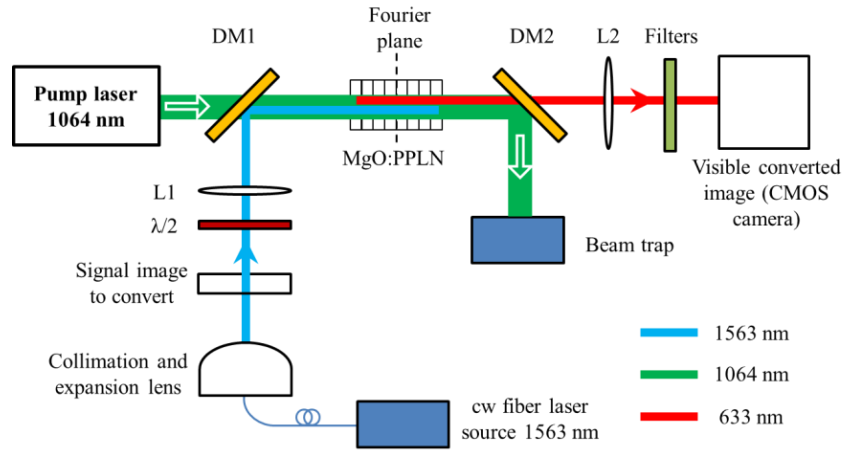

Fig. 3. Experimental setup used for upconversion imaging. L1 and L2 are lenses, DM1 and DM2 are dichroic mirrors.

1.5 mm x 1 mm. The upconverted wavelength is 633 nm. The crystal is positioned in the Fourier plane of a 4-f optical system where the focal length of L1 and L2 are $f_1 = 100$ mm and $f_2 = 125$ mm. A dichroic mirror (HT@633 nm, HR@1064 nm), a short-wave pass filter (HT below 750 nm) and two band pass filters (@635 nm, 10 nm FWHM and @633 nm, 5 nm FWHM) clean up the output image. The total transmission of the signal before conversion is 99% and the total transmission of the upconverted signal through the different filters is 89%. The upconverted image is recorded on a low noise CMOS camera with a quantum efficiency of 60% at 633 nm and a readout noise of 1.1 e$^-$ per pixel.

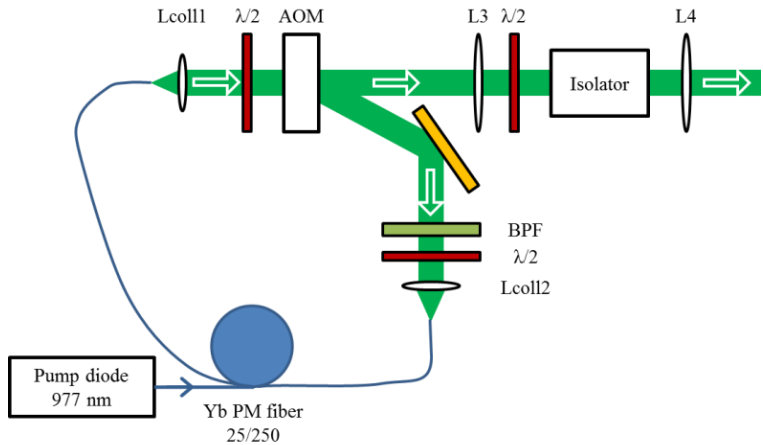

Fig. 4. Layout of the wide spectrum home-made laser. Lcoll1, Lcoll2, L3 and L4 are lenses and BPF a band-pass filter.

The first pump laser used is a commercial Q-switch YVO$_4$ laser (Spectra-Physics), with a 40 kHz pulse repetition rate, 22 ns pulse width, 7.5 W mean power and a narrow spectrum (<0.1 nm FWHM). The maximum peak pump power is 8.5 kW. A 600 mm focal lens fixes the pump beam diameter inside the crystal (1/e²) at 660 µm and the PPLN temperature is set to 97°C to satisfy the phase matching condition. The second pump laser, described in Fig. 4, is a home-made laser built to fulfil the desired spectral characteristics. A 3.5 m long, 25/250 µm core, polarization maintained ytterbium-doped fiber from Nufern is used as broadband gain medium. It is pumped by a 977 nm laser diode in contra-propagation. The free space cavity comprises two half-wave plates at the two ends of the fiber to control the polarization. An acousto-optic modulator (AOM) is used to Q-switch the laser. The cavity is closed on the first diffraction order of the AOM. A narrowband-pass filter (4 nm FWHM @1064 nm) is used to select the wavelength lasing range. An isolator and a half-wave plate enable us to control the power of the laser inside the upconversion crystal. The focal length of L4 is 125 mm and gives a pump beam diameter inside the crystal (1/e²) of 650 μm with a M² of 1.1 in both directions. At 40 kHz repetition rate, we obtain pulses of 250 ns with 3.2 W mean power, corresponding to a peak pump power of 320 W.

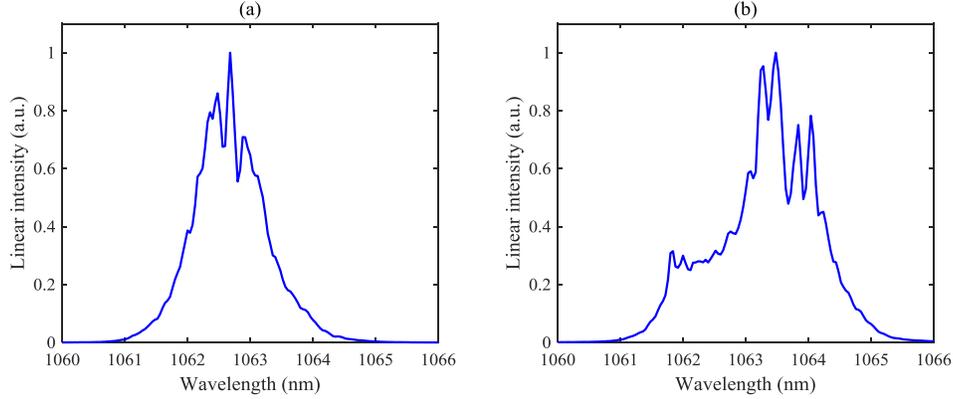

Fig. 5. Different spectra obtained with the home-made pump laser and used for imaging. The peak pump power is 320 W in both cases (a) 1 nm wide spectrum, (b) 2.7 nm wide spectrum.

Laser characteristics, in this second case, are very sensitive to AOM diffraction power and coupling inside the fiber. Thus, by changing the diffraction power, we control the lasing spectrum. Fig. 5 shows two examples of spectra obtained with our pump laser and used for imaging characterisation. The PPLN temperature is adjusted for each spectrum to reach the phase matching condition.

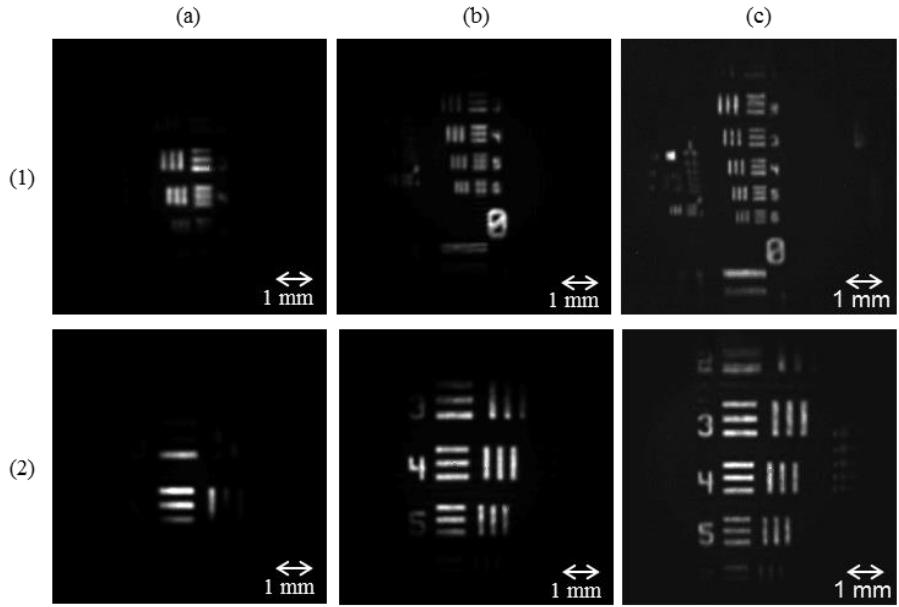

Fig. 6. Upconversion results. Lines (1) and (2) are two different positions on the resolution target. (a) Images obtained with the narrowband pump laser and 1.4 kW peak pump power. (b) Images obtained with the home-made pump laser with the spectrum of 1 nm wide and 320 W peak pump power. (c) Images obtained with the home-made pump laser with the spectrum of 2.7 nm and 320 W peak pump power.

A 1951 USAF resolution target is used to evaluate upconversion characteristics. Fig. 6 shows some images obtained with the two pumps described above. Images in line (1) are centered on the 1/5 element of the resolution target corresponding to 3.17 line pairs per mm (lp/mm) and images in line (2) are centred to the 0/4 element corresponding to 1.41 lp/mm. We observe a size factor between the image plane and the camera plane due to magnification of the 4-f optical system and the wavelength change [5]. The size factor $\gamma$ is given by:

$$\gamma = -\frac{f_2 \lambda_c}{f_1 \lambda_s} \sim -0.5 \qquad (10)$$

It can be measured using the resolution target calibration and the pixel pitch of the CMOS camera. Experimental observations agree with Eq. (10).

As expected by simulations, the field of view increases with the pump spectrum broadening. It is measured from upconverted images using the pixel pitch, the size factor $\gamma$ and the focal lens of L1. For the narrowband pump, the field of view of the system is 32 mrad whereas it is 70 mrad for the 1 nm wide spectrum and 111 mrad for the 2.7 nm wide spectrum.

## 4. Performance analysis

We present in this section the characteristics of our detection setup in terms of modulation transfer function, number of resolved spatial elements, conversion efficiency and sensitivity improvement compared to a direct detection with an InGaAs camera.

*4.1 Modulation Transfer Function*

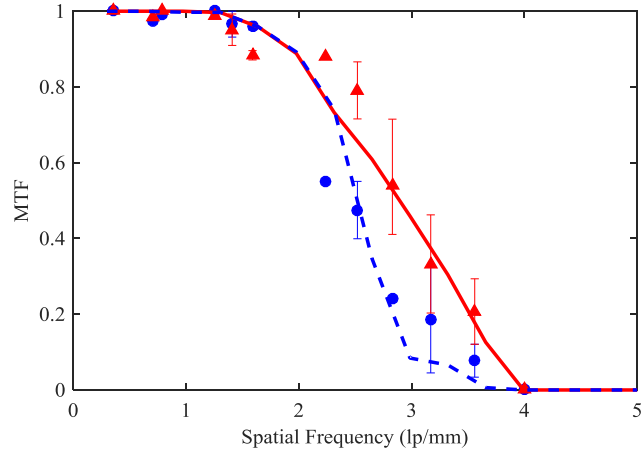

Fig. 7. Experimental MTF measured on upconverted images. Blue dots represent the MTF in the vertical direction. Red triangles represent the MTF in the horizontal direction. Lines represent the corresponding simulation results.

Fig. 7 represents the modulation transfer function (MTF) measured with different sets of images, the simulated MTF for the horizontal (plain red line) and vertical (dashed blue line) directions. There is a good agreement between simulations and experiments. Dispersion in experimental data is due to issues in alignment and measurements because of the high coherence of the illumination laser. Hence, the presence of speckle has a deleterious effect on images quality and for the estimation of the MTF. Resolution in the vertical direction is lower than in the horizontal one due to the smaller thickness of the crystal (1 mm against 1.5 mm). MTF is also uniform on the images. Moreover, the MTF is the same for the different pumps as long as the conversion is performed in the Fourier plane of the resolution target; the resolution is strongly dependent on the pump beam diameter that has the effect of a spatial frequency filter [12]. The resolution limit using the Airy criterion is MTF = 0.5 and is obtained with 2.5 lp/mm in vertical direction and 2.9 lp/mm in horizontal direction.

The corresponding number of spatial elements resolved for the different pump cases is summarized in Table 1. For the wider spectra, we obtain 64 spatial elements resolved in the horizontal direction, a more than threefold increase compared to the commercial narrowband pump.

**Table 1. Field of view and number of spatial elements resolved**

| Pump spectrum | Narrowband | 1 nm | 2.7 nm |
|---|---|---|---|
| Field of view (mrad) | 32 | 70 | 111 |
| Number of spatial elements in the vertical direction | 16 | 35 | 56 |
| Number of spatial elements in the horizontal direction | 19 | 41 | 64 |

*4.2 Conversion efficiency*

Conversion efficiency is a key parameter to characterize detection sensitivity. The photon conversion efficiency is measured using a 170 μm $1/e^2$ diameter Gaussian signal beam and the narrowband pump laser with a 660 μm $1/e^2$ beam diameter. To directly access the conversion efficiency inside the crystal, the depletion of the IR signal is recorded, at the maximum of the pump pulse, using an InGaAs photodiode. Fig. 8 represents the conversion efficiency as a function of pump peak power where error bars are lower than dots size. A maximum of 63 % conversion is achieved for a 1.2 kW pump peak power for a spatially single mode signal, with a collinear phase matching.

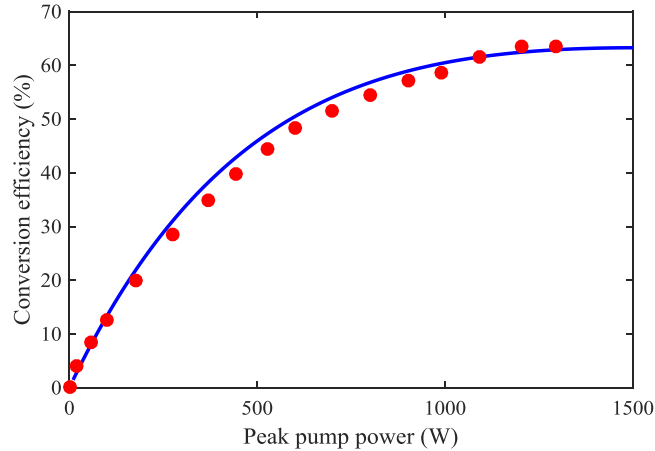

Fig. 8. Conversion efficiency as a function of peak pump power for a focalized gaussian signal beam. Red dots stand for experimental data and blue line for simulations.

The maximum conversion efficiency achievable using a longitudinally multimode pump laser has been explained in [21, 22]. It was shown to be 64 % for plane waves. The little discrepancy is due to the spatial Gaussian nature of the beams. Running our simulation code and including this theory, we obtained the blue plain curve without any adjustable parameter. Our data is in perfect agreement with simulations in the above case. We subsequently used the same code to access conversion efficiency in images.

Using a spatially uniform signal field in the object plane in numerical simulations, the influence of the non-collinear phase matching on the conversion efficiency is evaluated. Since the pump beam has a finite Gaussian size, the conversion efficiency decreases for larger phase-matched signal angles [23] due to a lower overlap between the pump and the signal. Fig. 9 represents the normalized conversion efficiency, as a function of the incoming signal angle on the crystal for different spectral widths of the pump. The pump for the dotted red curve is a narrowband 1064 nm laser and corresponds to the collinear phase matching condition of the system, with a field of view of 30 mrad. For the dashed blue curve, the spectrum is uniform from 1063.75 nm to 1064.75 nm and the corresponding field of view is 68 mrad. For the green curve, the spectrum is uniform from 1063.75 nm to 1066.45 nm and the obtained field of view is 128 mrad. Results are consistent with Eq. (5) and these figures are in very good agreement with experimental values mentioned above, except for the larger spectrum which is slightly narrower experimentally. This discrepancy can be attributable to a non-uniform pump spectrum. The conversion efficiency decreases moderately with non-collinear phase matching. Indeed, conversion efficiency is as high as 75 % of the maximum for a 125 mrad field of view. The field of view is ultimately limited by the physical crystal aperture to about 210 mrad in our experimental setup, since the thicker commercially available crystal for the considered poling period is 1 mm.

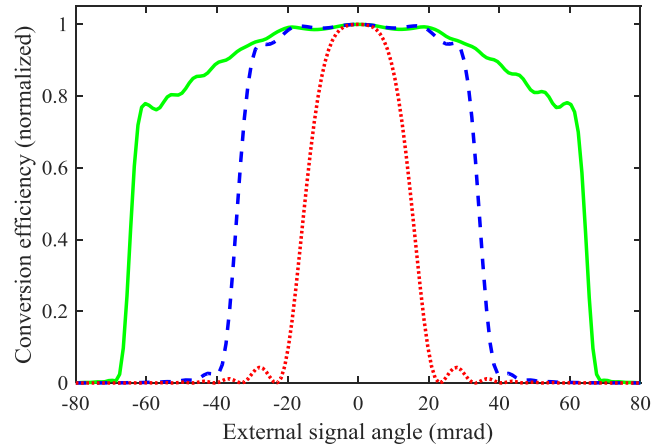

Fig. 9. Numerical results of the field of view of the upconverter for different spectral widths of the pump, where the conversion efficiency is given as a function of the incoming signal angle. In red dotted line, narrowband spectrum; in blue dashed line, 1 nm pump spectrum and in green solid line, 2.7 nm pump spectrum.

Performing simulations with a peak pump power set at 320 W and with the different pump spectra, we obtained 35 % conversion efficiency for the narrowband spectrum, 18 % conversion efficiency for the 1 nm wide spectrum and a 7 % conversion efficiency for the 2.7 nm spectrum. From these figures we can deduce an overall sensitivity for our system and compare it to a direct detection with a high performance InGaAs camera (80 % quantum efficiency and 30 e$^-$ readout noise). However, the conversion efficiency is considered for a Gaussian beam along the optical axis. As the pump beam has a Gaussian shape the conversion efficiency drops with the spatial frequency and the upconversion efficiency in images is slightly overestimated. Results are summarized in Table 2 where the overall efficiency takes into account CMOS camera quantum efficiency, total optical transmission and upconversion efficiency. Sensitivity improvement is calculated using the 1.1 e$^-$ readout noise of our CMOS camera, considered to be the dominant noise of our system.

As shown in Table 2, the pump spectral broadening for a fixed power has a detrimental effect on conversion efficiency for a given pump power. Fig. 10 represents the conversion efficiency as a function of the external signal angle on the crystal, for different pump spectra and a fixed 320 W peak pump power. Upconversion detection compares favorably with direct InGaAs detection. A factor 11 of sensitivity improvement is obtained with the narrowband pump laser which is not limited in peak power. Using the broadband pump laser, the same sensitivity than InGaAs camera is achieved, with a limited pump power. In order to obtain the maximum conversion efficiency of 63 % with the 2.7 nm wide spectrum, having more pump power is the one requirement; especially a peak pump power of about 9 kW is needed. In that case, the same sensitivity improvement than the one obtained with the narrowband pump spectrum and 1.2 kW peak power is expected. We were not able to experimentally obtain such a peak power with our laser. Further improvement in our pump laser will enable us to reach this power.

Table 2. Experimental conversion efficiency and sensitivity

| Pump spectrum | Narrowband | Narrowband | 1 nm | 2.7 nm |
|---|---|---|---|---|
| Peak pump power (W) | 1200 | 320 | 320 | 320 |
| Upconversion efficiency (%) | 63 | 35 | 18 | 7 |
| Overall efficiency (%) | 33 | 18 | 9.5 | 3.7 |
| Sensitivity improvement | x 11.4 | x 6.3 | x 3.2 | x 1.3 |

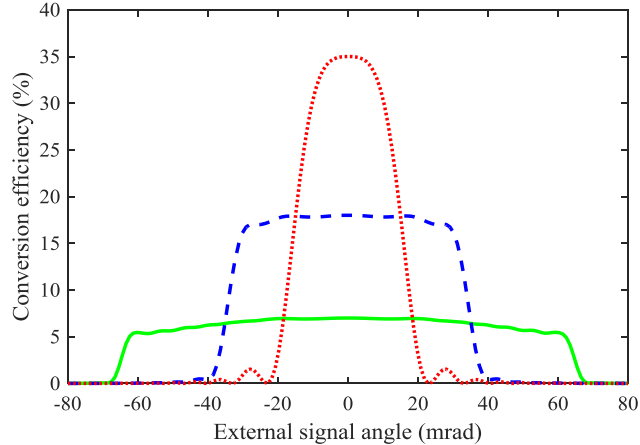

Fig. 10. Field of view and conversion efficiency for a 320 W peak pump power and different pump spectrum. In red dotted line, narrowband spectrum; in blue dashed line 1 nm pump spectrum and in green solid line 2.7 nm pump spectrum.

## 5. Conclusion

We have shown that by using a pump laser with a broadband spectrum we significantly increase the field of view of an upconversion imaging system. The number of resolved elements obtained is as high as 56x64 with a 2.7 nm wide pump spectrum, more than 3 times more in each direction than with a narrowband laser. We also demonstrate a conversion efficiency up to 63% using a narrowband pump laser, allowing a factor 11 of sensitivity improvement compared to a direct InGaAs detection. This sensitivity obtained with our broadband pump is close to the sensitivity that would be obtained with a direct detection. All our experimental results, image definition, MTF and conversion efficiency show a very good agreement with our numerical model. This work paves the way for practical use of upconversion detection for long range LIDAR and remote sensing applications where systems are limited by the sensitivity of detectors. Further work will focus on the development of a high peak power broadband pump laser to obtain both resolution and conversion efficiency.